# High Light-Efficiency Holographic Tomographic Volumetric Additive Manufacturing using a MEMS-based Phase-only Light Modulator

*Maria Isabel Álvarez-Castaño[1], Ye Pu[1], Christophe Moser[1]*

[1]*Laboratory of Applied Photonics Devices, School of Engineering, Ecole Polytechnique Fédérale de Lausanne, CH-1015, Lausanne, Switzerland*

**Abstract**

Light-based 3D printing, which relies on photocurable resins, has shown the capability to produce complex geometries with high resolution and fidelity. Tomographic Volumetric Additive Manufacturing (TVAM) employs a digital micromirror device (DMD) to project high-speed sequences of amplitude light patterns into a rotating resin volume, enabling rapid fabrication of 3D structures through photopolymerization. Typically, the light projection efficiency in such binary amplitude modulator-based systems is below a few percent. Recent advancements introduced phase encoding in TVAM using binary amplitude modulators, improving depth control and boosting light projection efficiency to approximately 10%. This was achieved by implementing the Lee hologram technique to encode phase into binary amplitude patterns. In this work, we present the first 3D printing platform utilizing a phase-only light modulator (PLM), based on an array of micro-electro-mechanical pistons. Compared to amplitude encoding, phase encoding with the PLM yields a 70-fold increase in laser power efficiency. By coupling this efficient light engine with a speckle reduction method in holographic volumetric additive manufacturing (HoloVAM), we experimentally demonstrate printing across different scales from hundreds of micrometers to centimeters using only digital control. The PLM opens up new avenues in volumetric AM for holographic techniques using low-cost single-mode UV laser diodes.

**Introduction**

Three-dimensional (3D) Additive Manufacturing (AM), better known as 3D printing, has been a breakthrough in many fields such as tissue engineering[1], regenerative medicine[2], aerospace[3], optical components, and many others. The first modern 3D printing method was light-based proposed by Kodama in 1981[4] which consisted of selectively solidifying material point-by-point or layer-by layer to build three-dimensional objects. Since then, various AM methods have been developed[5–8] using different materials. Layerless 3D technologies do not rely on layer-by-layer deposition. Such technologies are referred to as Volumetric Additive Manufacturing (VAM). Recently, several light-based VAM methods working with single–photon absorption have been developed. In reverse tomography, known as tomographic volumetric additive manufacturing (TVAM)[9,10], an entire three–dimensional object is simultaneously solidified after sequential amplitude light patterns are displayed into a rotating photoresin vial. In Xolography[11] and Light-Sheet 3D printing[12], two intersecting beams of different colors are required to perform the polymerization process, while dynamic light patterns are projected into the photosensitive resin.

Currently, most VAM techniques rely on amplitude patterns, commonly displayed using Digital Micromirror Devices (DMDs). These binary amplitude light modulators operate in reflection mode by tilting their micromirrors between two states: "on" and "off". In the "on" state, a micromirror directs light to illuminate a corresponding pixel or voxel on the printing plane, while in the "off" state light is directed elsewhere. DMDs generate grayscale patterns by rapidly toggling micromirrors between the 'on' and 'off' states, effectively controlling average light intensity. Recently, a holographic approach applied to tomographic VAM using phase encoding has been demonstrated[13]. This system uses Lee Holograms to enable the use of the binary DMD modulator as a fast phase modulator[14,15]. Using a binary amplitude modulator as a phase modulator has drawbacks such as poor light efficiency and pattern fidelity[13,14].

For decades, Liquid Crystal on Silicon (LCOS) Spatial Light Modulators (SLMs) were the only commercially available option for phase spatial light modulation. This device consists of electrically addressable pixels containing long chains of liquid crystal (LC) molecules positioned between two electrodes[16,17]. The alignment of LC molecules changes in response to an applied voltage, following the direction of the electric field. This realignment alters the intrinsic birefringence, resulting in phase and thus wavefront modulation[16]. Due to the LC molecule's viscosity, the standard frame rate range of the LCOS SLM is between 60 – 120 Hz. LCOS SLMs typically degrade when operated under UV light[18,19]. Recently, Texas Instruments introduced a new type of MEMS-based Phase Light Modulator (PLM) that provides phase retardation thanks to the vertical displacements of the mirrors in a piston fashion. The vertical motion of each micro-mirror can be independently addressed with a 4-bit displacement resolution (16 states or mirror levels). Currently, the evaluation module (EVM) offers frame rates of up to 1400 Hz and a fill factor of 95%, providing high speed and light efficiency. Since the PLM does not rely on LC molecules, it is less subject to pixel crosstalk, polarization-insensitive, and stable in phase as it has no molecular relaxations[20]. Due to these advantages, PLMs are gaining significant interest in applications such as wavefront shaping, holographic projection, and augmented reality displays[21,22].

In this work, we demonstrate the first VAM system implemented using the new MEMS phase-only modulator (i.e. PLM). We first measure and calibrate the 16 phase levels (4-bit) of the PLM using an interferometric method which uses a self-generated diffraction phase grating[23]

(see in Material and Methods session). We then measure the power efficiency of the holographic projections using the PLM and compare it with that of amplitude and phase encoding using a binary DMD. To leverage the PLM in TVAM, we introduce a new hologram generation pipeline for computing holographic projections with reduced speckles. For each projection angle, an axicon phase pattern is used to extend the depth of field in the reconstructed intensity pattern for a larger printing range along the direction of propagation after a Fourier lens. The pattern retention achieved with the axicon phase pattern is superior to a Gaussian beam profile over the diameter of the printing vial. To reduce speckles in the holographic intensity reconstruction, the reconstruction is digitally shifted laterally in a time sequence using nine distinct axicon phases, each with its vertex shifted to a different position, such that the averaging process during the printing greatly smooth out the speckles. Excellent print surface quality is achieved with a carefully chosen lateral displacement.

**Results**

The optical setup of our holographic VAM (HoloVAM) system using a Texas Instruments (TI) DLP67750 PLM is shown in Fig. 1a. The light source is a 405 nm single spatial mode laser diode. The PLM is positioned in front of a Fourier lens and a glass vial containing a photosensitive resin is placed at the conjugate plane (CP) of the focal plane of the lens L1(See the Methods section for a detailed description of the setup). During the printing process, the vial is rotated continuously at a speed determined by the PLM frame rate, and a sequence of pre-calculated full size holograms are projected to the resin. When High-Definition Multimedia Interface (HDMI) is used, the frame rate is 720 Hz and the vial rotation speed is 30 degrees per second. When Display Port (DP) is used, the frame rate is 1440 Hz and the rotation speed is 60 degrees per second. The PLM uses 16 discrete phase levels from 0 to $2\pi$ using 4 bits of data. The phase modulation is imparted by the micromirrors' vertical displacement as Fig 1b illustrates. During holographic reconstruction, a linear phase ramp is added to the hologram on the PLM to separate the reconstructed intensity pattern from the zeroth order by directing the desired first diffraction order to pass through a spatial filter (SF) for printing, which blocks all other diffraction orders, including the zeroth order. Details of the PLM operation and calibration method can be found in the Methods section.

Figure 1c shows a comparison of the pattern light efficiency among two amplitude and phase-encoded projections on a DMD (Table 1. left and center columns) and phase-encoded projections using a PLM (Table 1. Right column). We construct our setup (see details in Supplementary Fig. 1) in such a fashion that combines a DMD and a PLM so that experimental power efficiency can be compared.

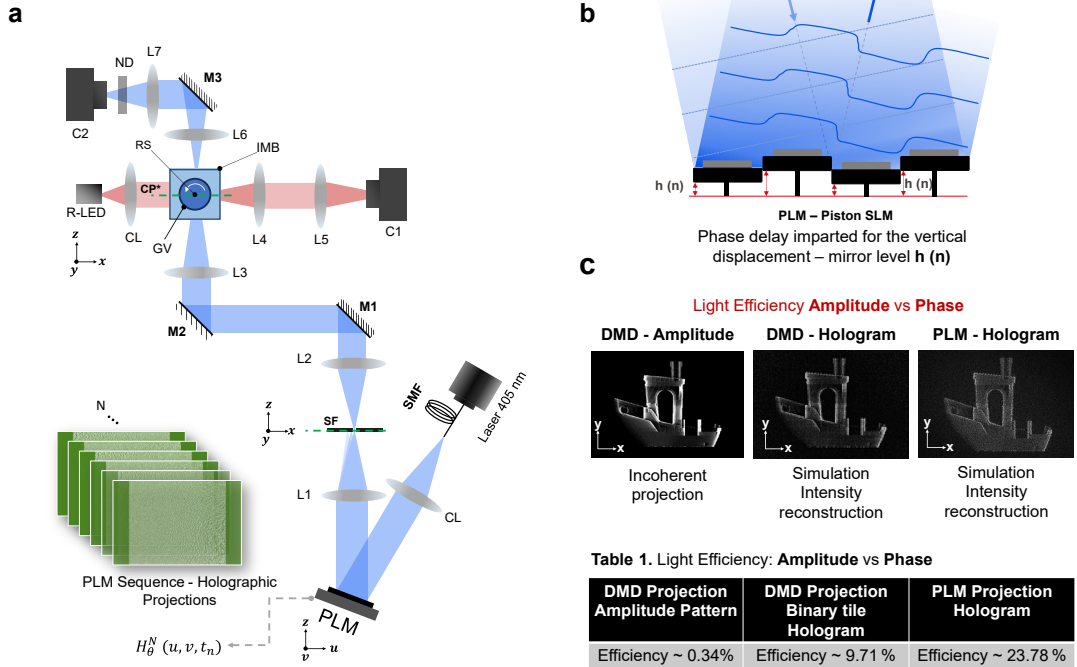

**Fig 1. Optical configuration and light efficiency. a.** Schematic of the optical configuration for HoloVAM using a PLM as a Spatial Light Modulator. An aspherical collimation lens (CL) collimates blue light from a laser diode coupled to a single-mode fiber (SMF). After the PLM, a Fourier lens (L1) is placed, with a spatial filter (SF) at the focal plane of L1 to remove undesired diffraction orders. A 4-f system (L2 and L3) conjugates the Fourier plane (CP*), which then serves as the printing region. A glass vial (GV) acts as the sample container and a rotation stage (RS) holds the sample container. An index matching bath (IMB) is used to minimize optical distortions caused by the curved walls of the cylindrical GV. **b.** Illustration of PLM row of micro-mirror pixels in piston fashion displacement to impart phase delay. **c**. Light efficiency comparison.

A blazed grating using a N-level phase-only modulator between 0 and $2\pi$ has a theoretical diffraction efficiency close to 100% (98.7%) in the first order of diffraction for N=16. Experimentally, we achieved a diffraction efficiency for a blazed grating close to 45%. This efficiency loss is due to the fill factor of the PLM (95%), reflectivity of the mirrors at 405 nm, and 20% light loss due to the PLM operation, in which the micromirrors are reset to zero position for a period of 140 µs every 694 µs within a frame. When a pattern is encoded, the absolute efficiency is close to 24%. With this latter efficiency value, the PLM light engine is approximately 70 times more efficient than amplitude projections and twice as efficient as DMD holographic encoding.

**Holographic Projections: A Computation Pipeline for Using a Phase a PLM**

The phase patterns were computed according to the pipeline in Fig 2. Our technique is based on the tomographic method, which assumes straight light rays. This assumption fails at small feature size due to diffraction. A recent theoretical study shows that above 30 $\mu m$ feature size, the straight light ray assumption is valid[24]. To obtain a uniform resolution throughout the print volume, the printable size is limited to the depth-of-focus of the beam. In our setup, for example, the resin container has an inner diameter of approximately $11\ mm$, which limits the printable feature to greater than 70 μm assuming a Gaussian beam. To effectively extend this depth-of-focus in the holographic projections, we modify the Point Spread Function (PSF) of the system using a Bessel beam. This approach produces a low-divergence beam in the printing region, which is a key parameter for achieving high and uniform resolution printing in a large printable size.

In our setup, the printing region is effectively located after the Fourier plane of the PLM. During the hologram pipeline construction process, we first generate an axicon phase $\varphi_{Ax}(u,v)$, which serves to generate low-divergence beams in the printing region, as shown in Fig. 2a. This axicon phase pattern is then added to the Computer Generated Hologram (CGH) phase pattern $h(u,v)$ retrieved from the intensity pattern of tomographic projections of the voxelized 3D model[1,10] using Gerchberg-Saxton (GS) algorithm. At the Fourier plane of the PLM, the reconstructed wave front is a convolution between the wave front of the CGH $h(u,v)$ and that of the axion phase $\varphi_{Ax}(u,v)$. After a short propagation distance beyond the Fourier plane, the reconstructed wavefront forms the desired projection image in which each bright pixel is a Bessel beam that stays focused over a distance much larger than the depth-of-focus of the same image without the axicon, as shown in Fig 2b (i) - (iv). Once the holographic projection pipeline is created, the stack of phase maps is sent to the PLM control software, which down-samples the phase to 4-bit, encode it properly into video frames for PLM control, and play the frames on the PLM screen. See PLM operation in the Methods section for more detail.

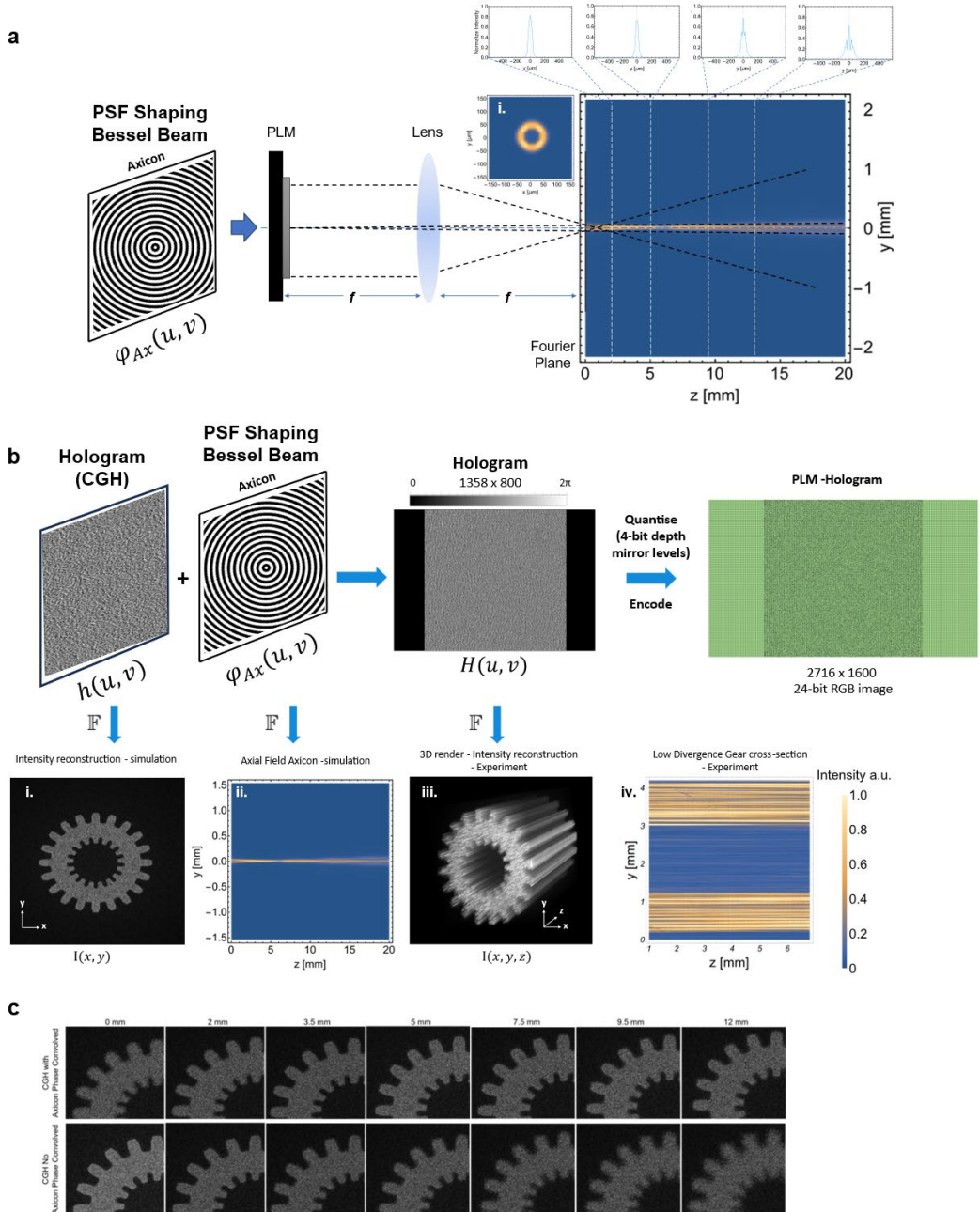

**Fig 2. Hologram Computation for printing with PLM. a.** Illustration of the working principle of the Bessel Beam generated with an axicon. **i)** shows the ring intensity distribution in the focal plane (Fourier Plane), which is characteristic of axicons. Profiles of the axial intensity propagation of the axicon are shown on the top where the axicon region starts after 2.5 $mm$ in the near field region. **b.** Pipeline for Computing the Holographic Projection $H(u,v)$. A CGH $h(u,v)$ computed by GS Algorithm using an amplitude tomographic projection as the target intensity is convolved with an Axicon phase $\varphi_{Ax}(u,v)$. **i)** Intensity reconstruction of a gear in the Fourier plane. **ii)** Intensity profile of the axicon along with the propagation direction. **iii) - iv)** shows a 3D rendering and cross section of the smeared information when the phase of the gear is convolved with the axicon.

## Speckle Evaluation and Reduction

Speckle refers to the high-contrast granular interference pattern commonly observed in optical reconstructions of Computer-Generated Holograms (CGHs) displayed on Spatial Light

Modulators (SLMs). Essentially, speckle is a result of the coherent superposition of a large number of wave fronts. In our system, the lack of amplitude modulation, the nature of imperfect phase retrieval from the GS algorithm, the pixelation of the PLM, defects in experimental conditions such as dusts and unwanted reflections all contribute to generate a speckle image in the printing region. Speckle in CGHs can be minimized using various techniques, including random phase methods, time multiplexing[24], tiling holograms[17,25,26]. Recent approaches also explore the use of Neural Networks for time-multiplexed speckle reduction[27].

Here, we use averaging of spatially shifted reconstructions to effectively reduce the speckle noise. This approach involves multiplexing up to $N_p = 9$ holograms per projection angle, each representing the holographic projection convolved with one of the 9 axicon phase laterally shifted off-vertex as illustrated in Fig 3a, b (see details in the method section). The result is a series of projections shifted around the axicon vertex position that, when played sequentially, reduces the speckle noise through the averaging effect in the photopolymerization during printing, which greatly enhances the quality of the printed objects. The optimal offset is determined by analyzing the speckle grain size using the power spectrum density (PSD)[28] of the image of the intensity reconstruction of projected holographic pattern[29], where the speckle grain size is $43.42\ \mu m$ (See Supplementary Note 2).

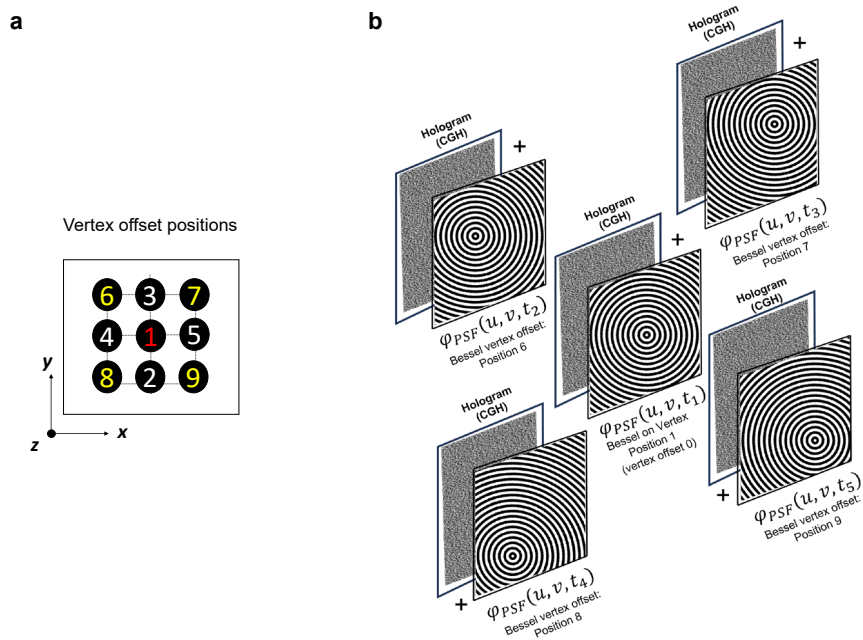

**Fig 3. Pipeline to generate a holographic projection with a reduced speckle. a.** Illustration of the 9 different vertex offset displacements in the lateral plane (XY) used to shift the image reconstruction a distance shorter than the speckle size. **b.** Example of the pipeline used to generate a reduced speckle noise projection.

Three different offset groups were used. A series of axicon phase patterns define each offset group shifted laterally from the central vertex in 9 different positions, as shown in Fig. 3a. When resulting holograms are displayed sequentially (time-multiplexed), statistically some peaks of the speckles will coincide with some valleys, reducing the contrast of the speckle intensity compared with a single projection.

The level of speckle reduction is analyzed with the speckle contrast coefficient, a commonly used metric to understand the speckle level of an image[28–30],

$$c = \frac{\sigma_I}{\langle I \rangle} \qquad (1)$$

where $\sigma_I$ is the standard deviation of the image intensity and $\langle I \rangle$ is the mean image intensity.

Figure 4a shows that the speckle contrast coefficient as a function of axicon displacement distance from $1\ \mu m$ to $33 \mu m$ with different number of holographic projections $N_p$, which reaches its minimum when the lateral displacements are approximately half the speckle grain size, at which the likelihood of the speckle peaks in one image coincide with the valleys in another is maximized. This trend is statistically confirmed in Fig 4b. Specifically, for $N_p = 5$, the minimum speckle contrast coefficient is $c = 0.45$, while for $N_p = 9$, the minimum possible value is $c = 0.33$. These results indicate that the optimal displacement distance, relative to the speckle grain size—effectively reduces speckle contrast. From Fig. 4c, we can observe that the intensity reconstruction appears smoother or less grainy when the displacement is close to half the speckle grain size, indicating that, for printing, we can improve the surface quality of the printed objects.

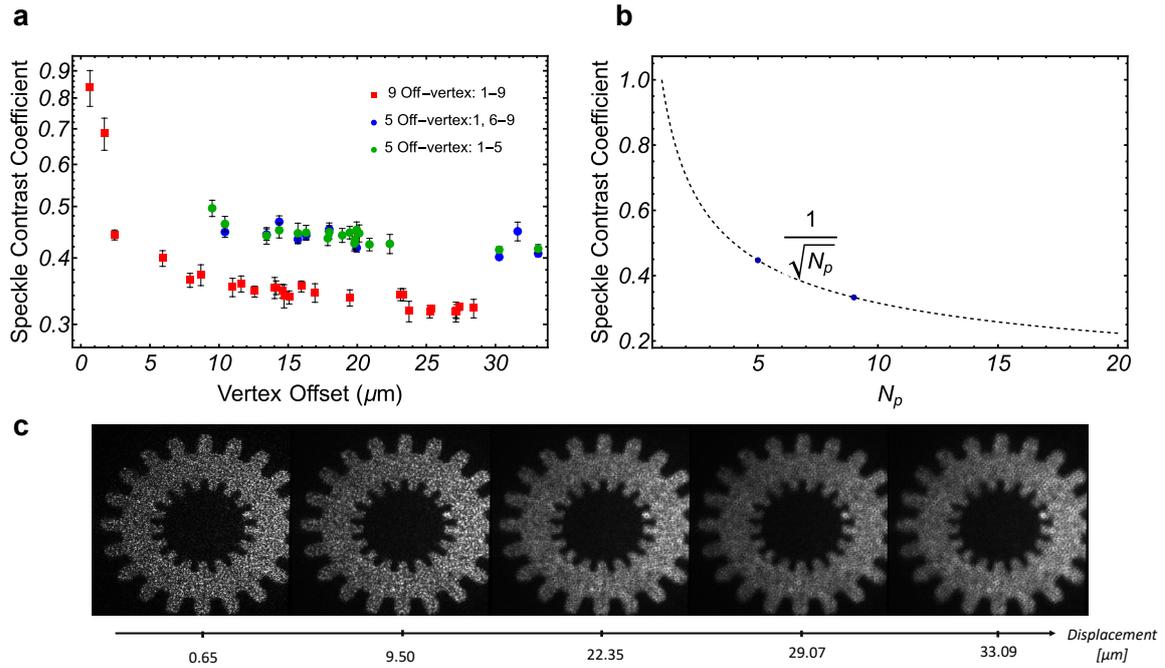

**Fig 4. Evaluation graphs for the speckle noise reduction technique. a** Measured speckle contrast coefficient graph versus the hologram lateral displacement. Error bars represent the standard deviation. Supplementary Fig. 3 shows histogram examples of the images used for the measurement. The lack of data points for the vertex offset between 24 μm and 30 μm is because we did not perform experiments at those positions. **b** Theoretical speckle contrast coefficient variation related to the number $N_p$ of hologram displayed.

**c** Examples of accumulated Intensity reconstructions of a gear when 9 different shifts on the axicon phase produce a lateral shift in the image plane. A plot of the profiles of the gear teeth from the images in 'c,' corresponding to different lateral displacements of the holograms, is shown in Supplementary Fig 4.

## Printing results

To demonstrate the capabilities of our PLM-based light engine for volumetric printing, we successfully printed multiple 3D models using holographic projections and the innovative speckle reduction approach described above. We used a commercial polyacrylate resin with Diphenyl (2,4,6-trimethylbenzoyl) phosphine oxide (TPO) as the photoinitiator (PI) at a concentration of 1 mM. Thanks to the efficient light engine, we produced objects at multiple scales through digital scaling of the holographic projection, as shown in Fig 5. Fig 5a shows a $4\ mm$ high fusilli 3D model printed in 32 seconds using a laser power of 18 mW. Fig 5b shows a large and small Stanford Bunny model. The large model is $8\ mm$ high and printed in 61 seconds with 50 mW laser power, while the small model is $4\ mm$ high and printed in 38.5 seconds using 20 mW laser power. Fig 5c shows a large and small model of a double helix DNA. The small model was printed in 23 seconds using 20 mW laser power. The zoomed inset of the large and the small models shows the good surface quality achieved with the speckle reduction technique.

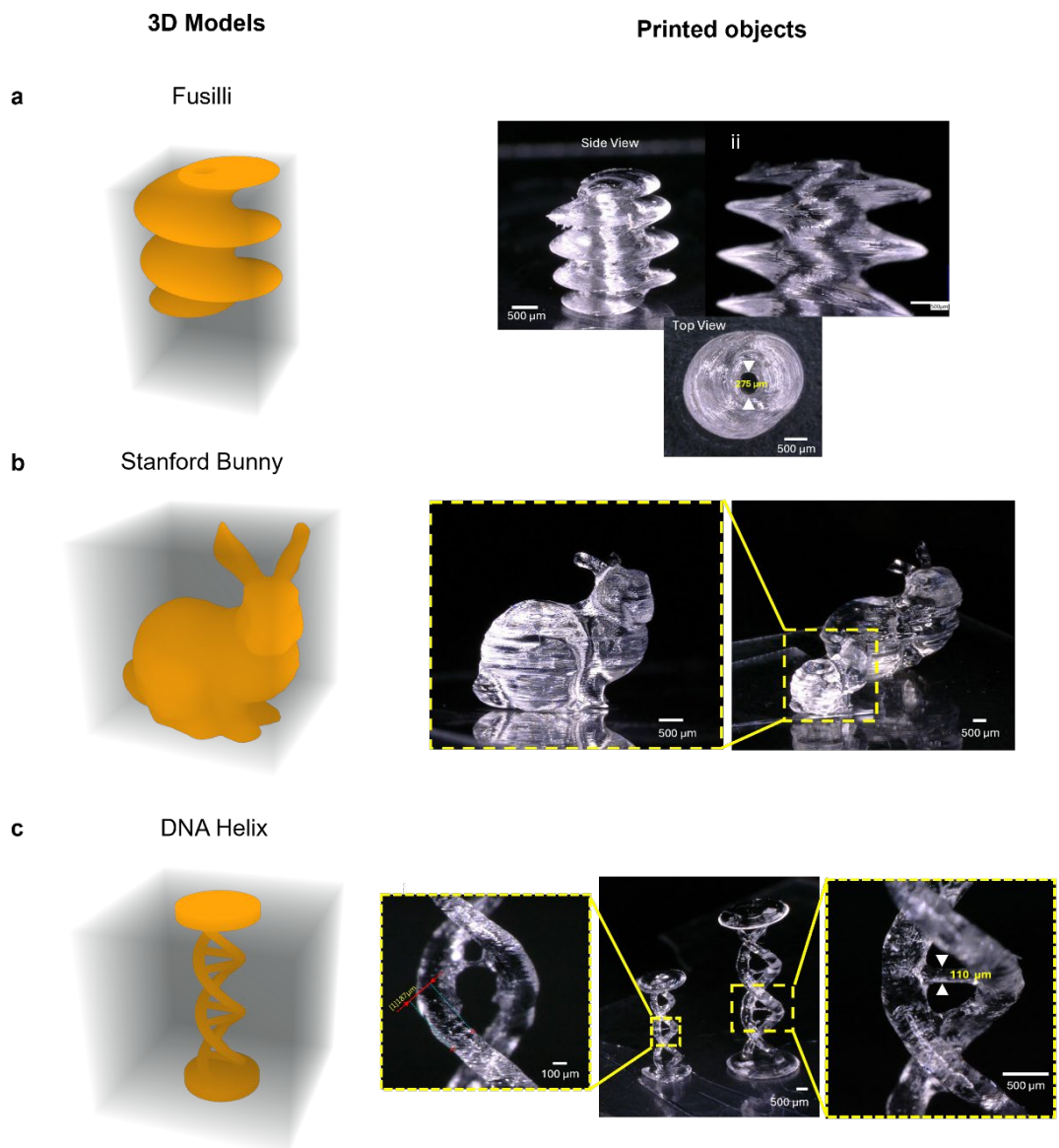

**Fig 5. Examples of printed objects with holographic VAM using PLM. a.** Left, 3D CAD model of a Fusilli. Right, printed part on acrylate-based resin of a Fusilli using the speckle noise reduction technique, Printing time: 32 seconds and 18 mW **b.** Left, 3D CAD model of the Stanford Bunny. Right, Big and Small models of a Stanford Bunny Printed with acrylate. The big model was printed in 61 seconds, using a laser power output of 50 mW. The small model was printed in 38.46 seconds, using a laser power output: 20 mW. **c.** Left, 3D CAD model of a DNA Helix. Right, Big and Small models of a DNA helix Printed with acrylate. Scale bars: $500\ \mu m$.

Figure 6 shows a DNA double helix printed using holograms without and with our speckle noise reduction technique. Fig 6a), Image of a printed part without the speckle noise reduction technique. Fig 6b) Image of the printed object printed with holograms with holograms with speckle noise reduction. Moreover, the granular pattern of the speckle introduces intensity gaps between the bright grains, which are visible in images captured by inspection camera 1. The application of the speckle noise reduction technique results in visibly smoother surfaces.

With reduced granularity in the holograms, there are fewer intensity gaps between the bright grains of the speckles, which in turn diminishes the formation of filaments, due to the self-focusing effect during the polymerization process, that can later cause delamination effects in the printed objects.

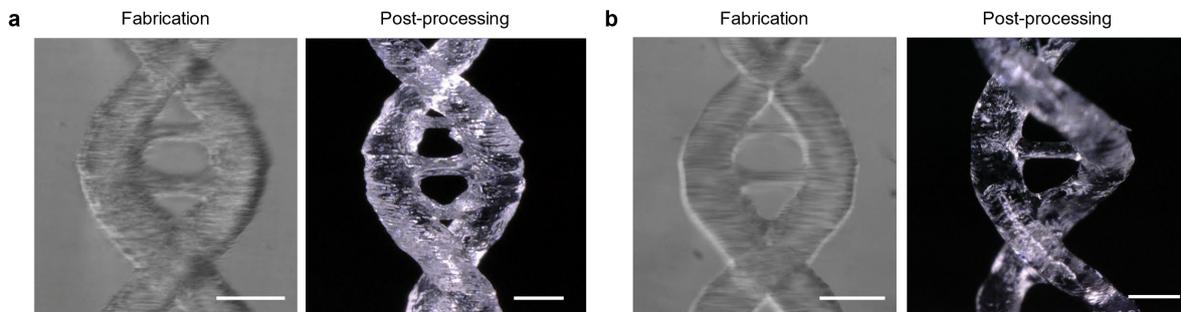

Fig 6. Surface quality improvement comparison. **a.** Zoom in of a DNA double helix printed without the speckle noise reduction technique. **b.** Zoom in of a DNA double helix printed with the speckle noise reduction technique. Images from inspection camera 1 (left) and post-processed images (right).

**Conclusion and Discussion**

In this work, we have demonstrated the potential of using the new MEMS-based, fast, phase-only modulator (PLM) in TVAM, which is capable of modulating light at a frame rate up to 1440 fps for holographic printing, marking a significant leap in holographic projection technology. By integrating the PLM, we achieved a more efficient light engine for HoloVAM, enabling faster print times and high-fidelity printed parts. Furthermore, our results show that holographic projections using the PLM are 58 times more efficient than amplitude-based projections, highlighting the impact of using phase encoding in VAM technologies. The elimination of tiling holograms and Lee holograms from the computational pipeline further simplifies the process while improving performance.

We have presented the working scheme of the software control for the DLP6750 PLM EVM, which is implemented in an in-house MATLAB program that quantizes the phase sequence to 4-bit, encodes it into the display frames for PLM control, and play the frame sequence at a constant rate without frame loss. We implemented a characterization method to measure and calibrate the phase response of the PLM. The correct characterization of the phase retardance with the corresponding voltage bias of a 4-bit phase allows us to improve the efficiency of the holographic reconstruction.

We also introduced a new computational pipeline to calculate the holographic projections. We also mitigated the effect of the speckle by using a time-division multiplexing method where a series of grouped CGHs per angle convolved with up to 9 different PSF off vertex allow us to improve the light dose while reducing the speckle noise.

**Materials and Methods**

*Experimental Setup for 3D printing*

The schematic of the optical setup for holographic tomographic additive manufacturing (HoloVAM) using a PLM as a spatial light modulator is shown in Fig 1a. A fiber-coupled continuous wave (CW) 405 nm blue laser diode (Integrated Optics, 0405L-13A-NI-AT-NF) is used as the light source. This beam is collimated using an aspherical lens (CL) and then obliquely incident upon the PLM arranged in a Fourier configuration with the Fourier lens L1 (focal length f1 = 180 mm), and a spatial filter (SF) is used to filter out the zero-order diffraction at the Fourier plane of L1. Lenses L2 (f2 = 150 mm) and L3 (f3 = 200 mm) form a 4-f system that conjugates the Fourier plane and rescales the holograms by 1.33X in the printing plane (CP*, Conjugate Fourier Plane). An index matching bath (IMB) is used .

A rotary stage (RS, Zaber- RSW60C-E03T7-KX13A) holds the sample vial (GV, glass vial), which is set to rotate at a constant speed of 30°/s when the PLM is using HDMI or 60°/s when DP is used. Because of the two different PLM frame rate possible settings, a different time per turn was set when the PLM interface was changed; when using the HDMI interface, the time per rotation is 12 seconds, and when using DP interface, the time per rotation is 6 seconds. Using the in-house MATLAB software, the PLM displays a sequence of holographic projections every $\Delta\theta = 0.5°$ at a frame rate of 720 Hz (HDMI) or 1440 Hz (DP), corresponding to light dose time per angle of $\sim 16\ ms$ and $\sim 8\ ms$, respectively. The Trigger 2 output of the PLM is used as a counter to ensure precise synchronization between the rotation platform and the hologram sequence. A computer laser control program and a mechanical shutter are used to allow the laser to irradiate the resin sample at a specified power and for a given period. An index-matching bath of vegetable oil (n = 1.48) is used to mitigate the lensing effect caused by the cylindrical vials containing the photoresin.

Two inspection systems were incorporated to monitor the polymerization process and the holographic projections. The first system use a Red LED (R-LED) collimated using a collimation lens (CL) (polymerization monitor) employs lenses L4 (f4 = 100 mm) and L5 (f5 = 180 mm) with camera 1 (C1, iDS UI307xCP-M), while the second system (projections monitor) uses lenses L6 (f6 = 75 mm) and L7 (f7 = 100 mm) with camera 2 (C2, iDS UI327xCP).

*PLM operation*

The Texas Instruments PLM is a phase-only SLM based on vertically moving micromirrors. The DLP6750 PLM EVM used in this work features a 1348 x 800 pixel array with a pixel pitch of 10.8 $\mu m$ and uses a video interface (HDMI or DP) for data communication with the host computer. Figure 7 shows the operational principle of the PLM micromirror. The mechanical structure of the PLM pixel is shown in Fig 7a. The micromirror (grey) sits on the metal hinge (blue), which is maintained at the voltage of the bias voltage electrode (EB) on the substrate. Four control electrodes (E0 – E3) can be switched on or off to a control voltage. The potential difference between the control and the bias electrodes provides an electrostatic attraction force that is programmable depending on the number of control electrodes that are switched to the on state, causing the movement of the micromirror (red arrow). Consequently, the position of the micromirror modulates the phase of the reflected light. Figure 7b shows the structure of the electrodes and their driving circuitry. The micromirror actuation is controlled by four memory cells that are each connected to one of the control electrodes, the content of which is distributed from a computer through either a display interface or a universal serial bus. Owing to the memory cells, the PLM possess an important advantage of no flickering over liquid-crystal based SLMs. Furthermore, the PLM device is polarization insensitive, and adapting it to work with different wavelengths can be simply achieved through adjusting the bias voltage without the need for recalibration.

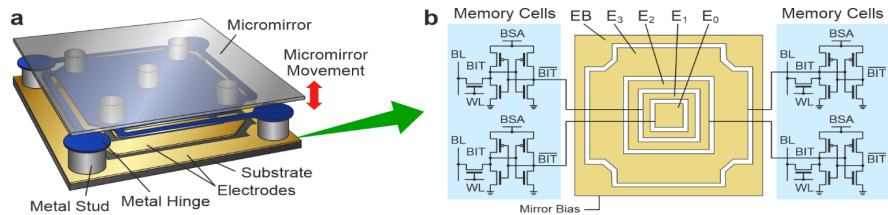

**Fig 7**. **Operational principle of the Texas Instruments' MEMS-based PLM. a.** Illustration of the mechanical structure of the PLM pixel adapted from the PLM User Guide. **b**. Structure of the electrodes and their driving circuitry. Note the inversed bits for E0 and E1. Bit lines (BL), word lines (WL). Note that E0 and E1 bit are inverted

Figure 8 shows the working scheme of the control software for the DLP6750 PLM EVM. The architectural layering of the control software is shown in Fig 8 a. Once connected to a host computer through HDMI or DP, the PLM functions as an additional display screen of 2716×1600 pixels on the computer side after basic configurations using a USB-based control software. The bit mapping from the computer display frame to the PLM EVM is illustrated in Fig 8 b. The actuation of each PLM pixel is controlled by four pixel bits from a 2 x 2 pixel block in the corresponding 24-bit RGB display frame received from the computer video interface.

This allows the PLM to interpret each received display frame data at 2716×1600 pixel resolution as 24 frames of 4-bit micromirror positions at 1358×800 pixels per frame, achieving a PLM frame rate of 30×24 = 720 Hz and 60×24 = 1440 Hz with HDMI (30 fps) and DP (60 fps) interface, respectively. The task of the control program is to reverse-encoding from the phase map required to be displayed on the PLM to the computer side so that each phase map sequence can be correctly displayed on the PLM without frame loss. While the GPU-accelerated encoding step is completed offline, playing the encoded frames on the display adapter at the video rate must be performed in real time. While there is no guarantee for response time in a high-level programming language like MATLAB or Python and a time-sharing operation system such as Windows, we achieved nearly loss-free playing of phase maps owing to the support from Psychtoolbox-3 and the double-buffering capability of OpenGL without resorting to C/C++ program and DirectX.

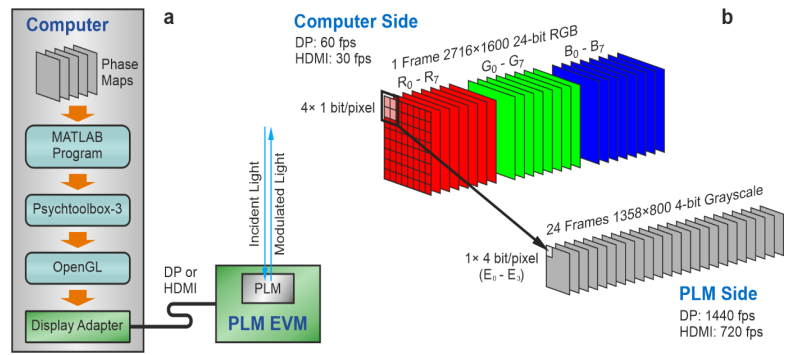

**Fig 8**. Working scheme of the control software for the DLP6750 PLM EVM. **a** Architectural layering of the control software. An in-house MATLAB program receives the phase sequence to be displayed on the PLM, resamples it to 4-bit, and encode it into computer display frames. Under the assistance of the PsychToolbox-3 package and OpenGL in the operating system, the display frames are sent to the PLM EVM through the display adapter and the DP or HDMI cable. **b**. Bit mapping from the computer display frame to the PLM. From the computer perspective, each display frame contains a bitmap of 2716×1600 pixels, each of which is 24-bit red-green-blue (RGB) data encoded in 8 bit per color. From the PLM perspective, on the other hand, each phase map contains 1358×800 pixels, each of which is 4-bit grayscale data providing 16-level phase modulation. The control electronics in the PLM EVM treats each incoming display frame from the computer as 24-planes of one-bit bitmap and bins 2×2 pixels of each plane into one 4-bit pixel data (E0 – E3) for PLM phase control. The PLM EVM can receive video data at 30 fps through HDMI or 60 fps through DP. Therefore, during the time of one video frame, 24 phase maps are displayed on the PLM, achieving 720 Hz (HDMI) or 1440 Hz (DP) of phase modulation rate.

*PLM Calibration*

The PLM is capable of producing a phase shift of up to $2\pi$ for a wavelength range of $405\ nm < \lambda < 650\ nm$, which is wavelength-dependent. The PLM provides a mirror bias voltage control that is used to adjust the mirror displacement according to the wavelength. The calibration is achieved by measuring this phase delay as a function of the micro-mirror state.

A calibration was performed for the correct operation of PLM. Several phase calibration methods for phase-only modulators based on LCOS SLM have been reported in the literature. Common methods include phase shift interferometry based on the analysis of interference fringes obtained in a Michelson interferometer[16], two-beam interferometry[16,21], binary diffraction grating, and many others based on the Muller matrix and Jones vectors[31–33]. In contrast to LCOS, where the phase delay is based on birefringence and each pixel is a Pockels cell, the operation of the PLM is based on the phase delay imparted for vertical displacements of the micromirror. Therefore, the phase delay in each pixel is a function of the 4-bit response of the mirror position. Our calibration is an adaptation of an interferometric method that uses a self-generated diffraction grating[23]. Fig 9a shows the optical setup used for phase calibration. For this method, the image displayed on the PLM consists of two sectors, as shown in Fig 9b. One part of the image consists of a binary grating, which is kept constant, while the other part of the image is a uniform gray level that varies. These gray levels correspond to different mirror positions. When a collimated beam is incident on the PLM, the binary grating region diffracts the light, splitting the beam into several diffraction orders, while the uniform gray level region (piston) modifies the optical path of the reflected light. Consequently, the shifts in the resulting interference fringes between the diffracted and the reflected beams is proportional to the induced phase change in the reflected beam, which is used in our calibration.

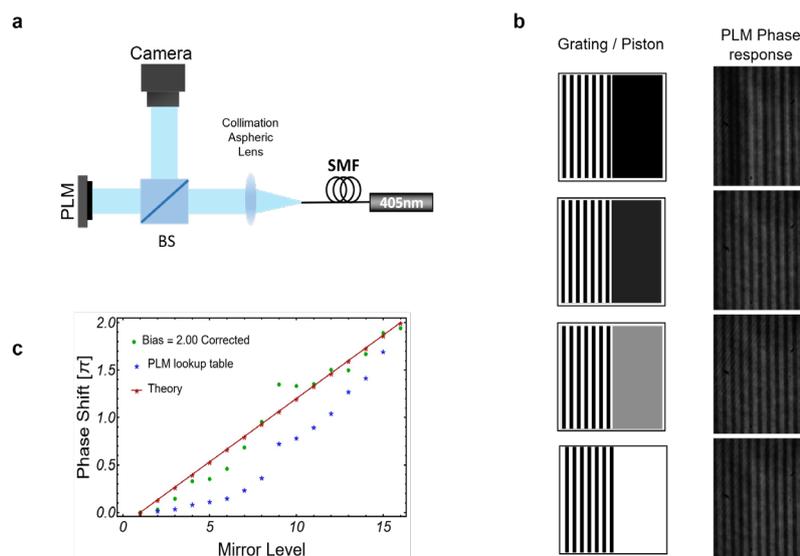

**Fig 9. Phase calibration. a.** Phase calibration diagram **b.** (Left row) Phase is displayed on the PLM for calibration. (Right row) corresponding fringe patterns. **c.** Phase modulation graph relative to the mirror level. Blue represents the lookup table provided by the manufacturer, the red represents the desired linear performance of the phase delay.

*Time Multiplexing of Computer-Generated Holograms with Lateral Shift*

The lateral shift imparted to the holographic projections is achieved by convolving the CGH phase with axicon phases with different offsets. When these holographic projections are displayed sequentially over time, the accumulated intensity generates a holographic projection with reduced speckle noise, resulting in a smoother image as explained above. However, a small lateral shift produced by an axicon with its vertex off-axis can generate a slight angular divergence in the light intensity distribution during near-field propagation, which is also transferred to the intensity reconstruction of the desired shape when the holographic projection is convolved with an axicon off-axis (See Supplementary Information Note 4. Supplementary Fig 5.a-b, angular shift of the axicon axil intensity).

To evaluate the shift imparted to the holographic projection, we determined the intensity centroid position of the reconstructed gear pattern along the propagation distance. The Intensity Centroid is a spatial intensity measurement that provides the coordinates of the intensity-weighted centroid, which we then track over the propagation distance. Fig. 10a illustrates two examples of the shifts imparted to reduce speckle noise. The vertex shift generates a lateral displacement of the holographic projection in the y-direction of approximately $\sim 12\ \mu m$ (Fig.10a -top) and $\sim 31\ \mu m$ (Fig.10a -bottom). In both cases, a slight angular divergence of $\pm 3\ \mu m$ is noticeable along the propagation. However, when the time multiplexing of the holographic projections is performed, the centroid exhibits an average position that propagates parallel to the axicon on-axis, with a divergence of less than $3\ \mu m$. The small fluctuations in the intensity centroid are due to changes in the speckle pattern during propagation, which are influenced by the Bessel intensity distribution of the point spread function (PSF). This effect enables us to achieve prints with smooth surface quality, similar to previous works such as blurred tomographic printing[34].

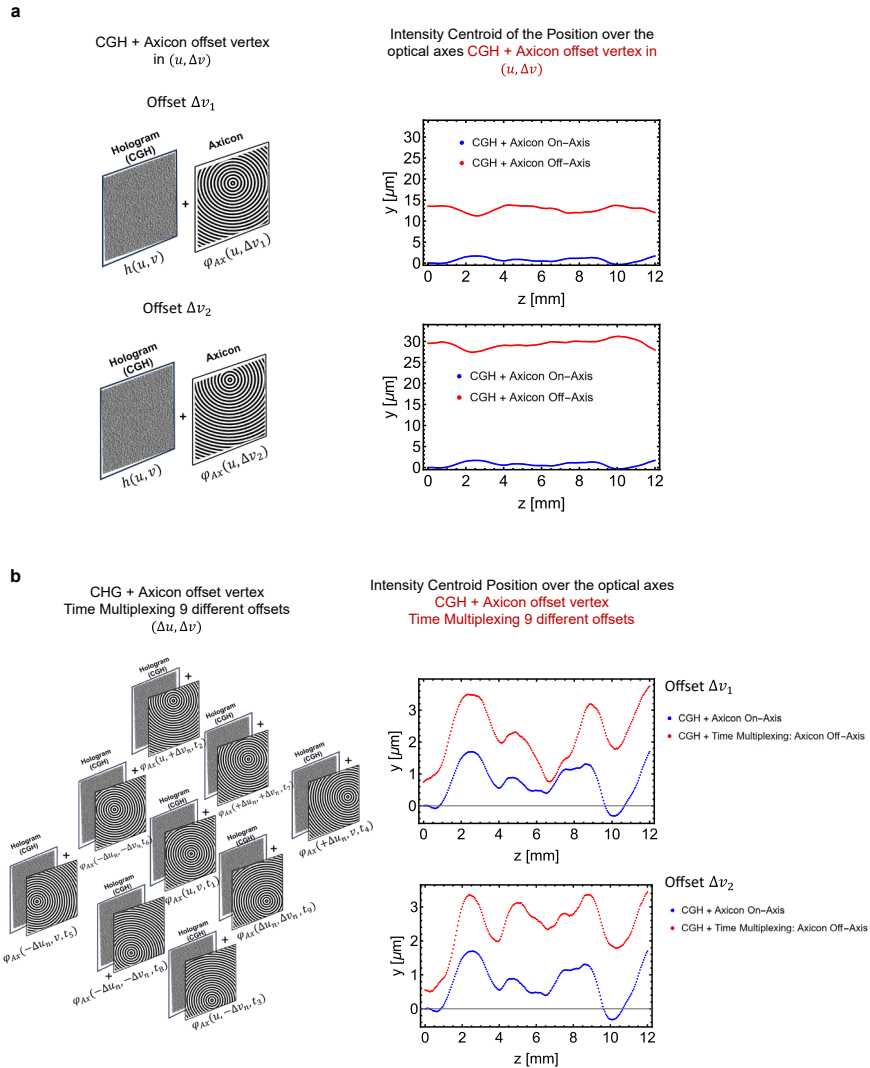

**Fig 10. Phase hologram used to generate holographic projections with a lateral shift. a** Left: Phase hologram of the off-axis axicon convolved with the calculated computer-generated hologram (CGH) of a gear. Right: Plot of the gear intensity centroid position over the axial propagation. **b** Left: Calculated computer-generated hologram (CGH) of a gear convolved with the 9 Phase hologram of the off-axis axicons. Right: Plot of the averaged gear intensity centroid position when the CGH is multiplexed with Axicon offset in 9 different lateral positions.

*Photoresin*

Photoresin was formulated by combining the photoinitiator TPO (Diphenyl (2,4,6-trimethylbenzoyl)-phosphine oxide, Sigma) with a commercial polyacrylate photoresin (PRO 21905, Sartomer) at a concentration of 1 mM. This mixture was homogenized using a planetary mixer deaerator (Kurabo Mazerustar KK-250SE). The photocurable resin was subsequently transferred into cylindrical glass vials (12 mm outer diameter) and subjected to sonication to eliminate air bubbles.

*Post-processing*

Once printing was complete, the printed parts were extracted from the glass cylinders and immersed in propylene glycol monomethyl ether acetate (PMGEA) for a 10-minute rinse under gentle stirring using a vortex mixer. Subsequently, they underwent an additional 10-minute cleaning process in isopropyl alcohol (IPA). The printed samples were post-cured under UV for 10 minutes.

*Photography*

The printed parts were analyzed using a Keyence digital microscope (VHX-5000) with magnifications ranging from 20x to 200x.

*Simulations*

3D renderings of .stl files, along with recorded light intensities and simulated intensity distributions, were generated using Wolfram Mathematica 13.1[35]. The tomographic projections were calculated through the Radon transform (Intensity targets for the holographic projections), as well as CGHs generation were calculated with MATLAB[36][33] and Wolfram Mathematica 13.1.

### Acknowledgments

This project has received funding from the Eurostars-3 (VOLTA-E!3908) joint program with co-funding from the European Union's Horizon Europe research and innovation program, Innosuisse (Swiss Innovation Agency), and Innovation Fund Denmark (IFD).

### Author Contributions

M.I.A.C.: conceptualization, experiments, data acquisition, interpretation, and validation, simulations. Writing: original draft, reviewing, and editing. Y. P.: PLM control software development, co-supervision. Writing: original draft, reviewing, and editing. C.M.: conceptualization, methodology, supervision. Writing: original draft, reviewing, and editing.

### Competing Interests

CM is a shareholder of Readily3D (Switzerland), a company that develops and commercializes tomographic volumetric 3D printers. The authors declare no other conflicts of interest.

***Data availability.***

The data presented in this paper are available upon reasonable request.

# Supplementary Information

**High Light-Efficiency Holographic Tomographic Volumetric Additive Manufacturing using a Phase-only Light Modulator (PLM)**

*Maria I. Álvarez-Castaño[1], Ye Pu[1], Christophe Moser[1]*

[1]*Laboratory of Applied Photonics Devices, School of Engineering, Ecole Polytechnique Fédérale de Lausanne, CH-1015, Lausanne, Switzerland*

**Supplementary Note 1: Light efficiency measurements:**

Supplementary Figure 1 shows the experimental setup used for the volumetric 3D printer based on holographic projections. The setup integrates two light engines: one using a DMD with binary holograms, and another using a PLM with 4-bit holograms. In the previous work by Álvarez-Castaño et. al.[13], the efficiency was measured along the light path indicated in light blue. The amplitude and phase efficiencies for the DMD were reported ass $\eta_{\text{phase}}^{\text{DMD}} = 0.34\%$ and $\eta_{\text{phase}}^{\text{DMD}} = 9.71\%$, respectively. The measured power corresponds to the amplitude projection and the holographic reconstruction of the same target shape, with the same expected reconstruction size. Similarly, we measured the input power (light incident on the PLM), which was $P_{\text{in}} = 38.03 \text{ mW}$. A computer-generated hologram (CGH) producing the same intensity reconstruction was then displayed on the PLM. Subsequently, the power was measured in the conjugate plane of the Fourier plane (CP*). The output power after the 4F system (printing plane) was $P_{\text{out}} = 9.04 \text{ mW}$, resulting in a pattern efficiency of $\eta_{\text{phase}}^{\text{PLM}} = 23.78\%$.

A diffraction grating with a linear carrier of $\Lambda = 4$ where displayed on the PLM, which is the linear phase added to the holographic projections to filter out the zero order of diffraction from the patterns in the printing plane. this measurements give us $45.91\%$

$$\frac{I_{=+1}}{I_0} = \frac{2.53 \; mW}{5.51 \; mW} \tag{S-0}$$

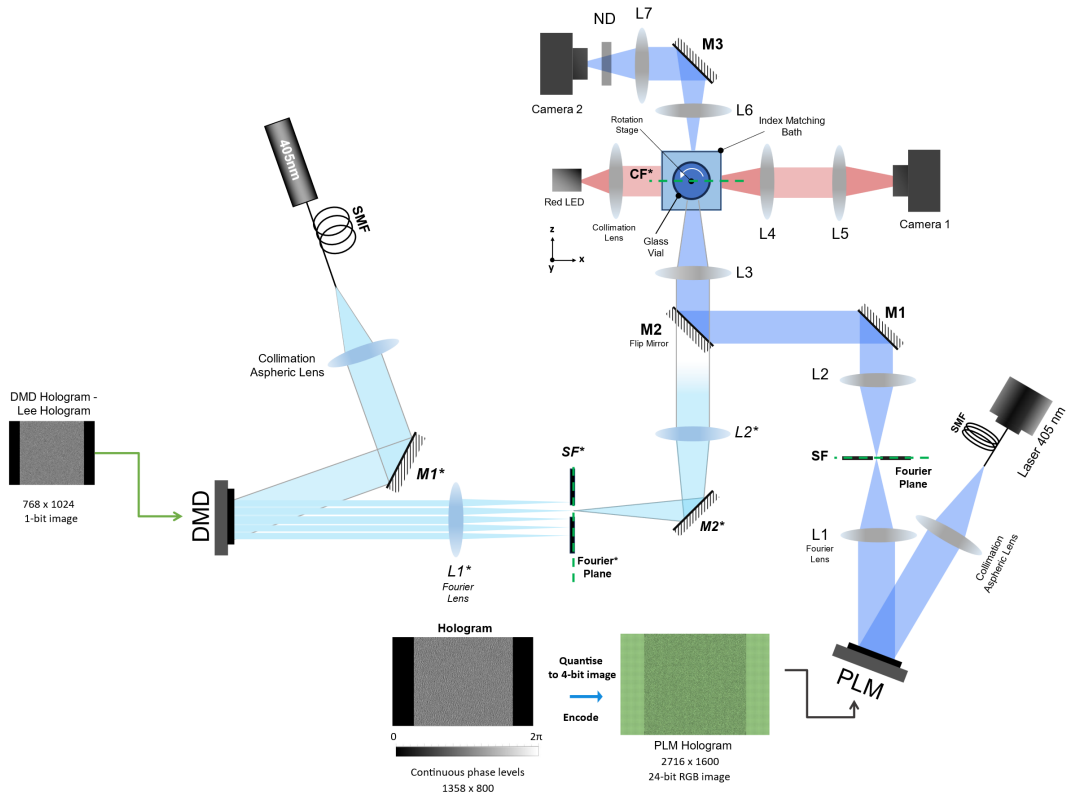

**Supplementary Fig. 1. Experimental setup of the Holographic Tomographic Volumetric Additive manufacturing (HoloVAM) using two different light engines.** The light blue color indicates the light path trajectory of the light engine that uses a DMD (Vialux DLP7000, 1024 × 768 resolution, pixel size 13.76 μm) as a spatial light modulator. In this configuration, mirror M2 is flipped to direct the light toward the printing plane (CP*, Conjugate plane), which is the conjugate plane of the Fourier plane of lens L1*, formed using lenses L2* and L3 (Green dashed lines). A Spatial Filter (SF*) is placed in the Fourier plane of the L1* to allow the propagation of the order -1 which is the most efficient when using the Lee Hologram method. The darker blue color represents the light path of the light engine using a PLM (TI DLP6750 PLM EVM, 1358 × 800 resolution, pixel size 10.8 μm). For this path, the printing plane is also the Conjugate Plane (CP) of the Fourier lens L1 (green dashed lines), using the lenses L2 and L3. A Spatial Filter (SF).

## Supplementary Note 2: Power Spectral Density

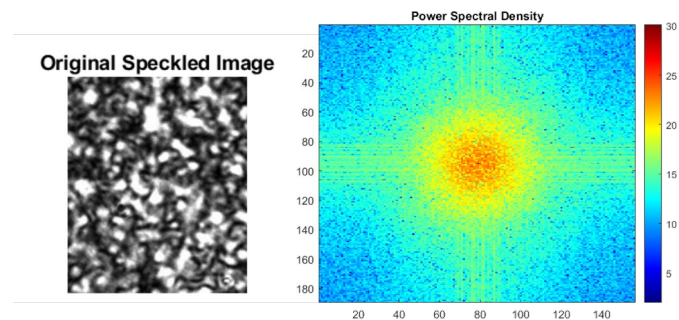

**Supplementary Fig. 2.** Left, original speckle image. Right, Power Spectral Density

To implement the method, it is necessary to calculate the appropriate lateral displacement meaning the off-vertex shift required to reduce the speckle noise across the multiplexed holograms. The optimal offset is determined by analyzing the speckle grain size using the power spectrum density (PSD) of the image of the intensity reconstruction of projected holographic pattern (See Supplementary Note 2). To obtain the power spectrum $P(u,v)$ we use the equation (1)[25].

$$P(u,v) = \frac{|FFT\{I(x,y)\}|^2}{L*W} \tag{S-1}$$

Where, FFT stands for Fast Fourier Transform, and $L$ and $W$ represent the vertical and horizontal components of the image. Here, the inverse of the dominant spatial frequency corresponds to the average speckle grain size, which in our measurement is $43.42\ \mu m$. The speckle grain size in our technique corresponds to the vertex displacement. By accumulating the intensity reconstruction over time from $N_p$ holographic projections that are laterally displaced with 'uncorrelated' speckle noise, due to shifts imparted by the PSF (axicon out of the vertex), we expect the speckle contrast to decrease.

## Supplementary Note 3: Speckle contrast coefficient measurement

We collected images from the experimental setup within the near-field region to obtain the intensity reconstruction of a gear. During the experiment, we time-multiplexed the CGH phase convolved with nine different axicon phases, which produced different lateral displacements. All images were acquired using the same integration time, with the PLM operating at a frame rate of 1440 fps. To measure the speckle contrast coefficient in the experimental intensity reconstructions, we analyzed different patches in the images, as shown in Supplementary Fig. 3 (top row), and calculated the speckle contrast coefficient using equation (1).

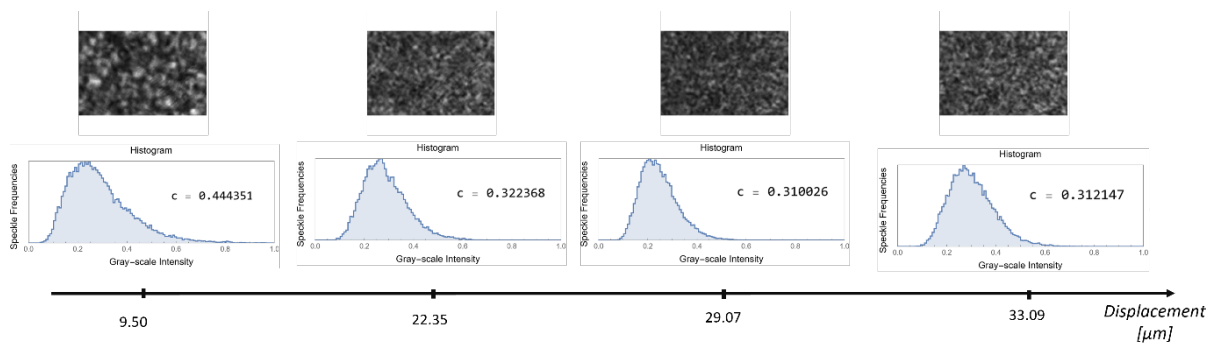

**Supplementary Fig. 3. Speckle Image Analysis.** Top row: An example of one of the patches analyzed from a single image. Each column represents a different displacement. Bottom row: Histogram distribution of the images with their corresponding speckle contrast coefficients.

As the lateral displacement increases, the gap between the maximum and minimum values of the bright grains decreases (see gear teeth profile in Supplementary Fig 4), which improves the light dose during the printing process, this result will improve the surface quality and which prevents the printed objects from delaminating.

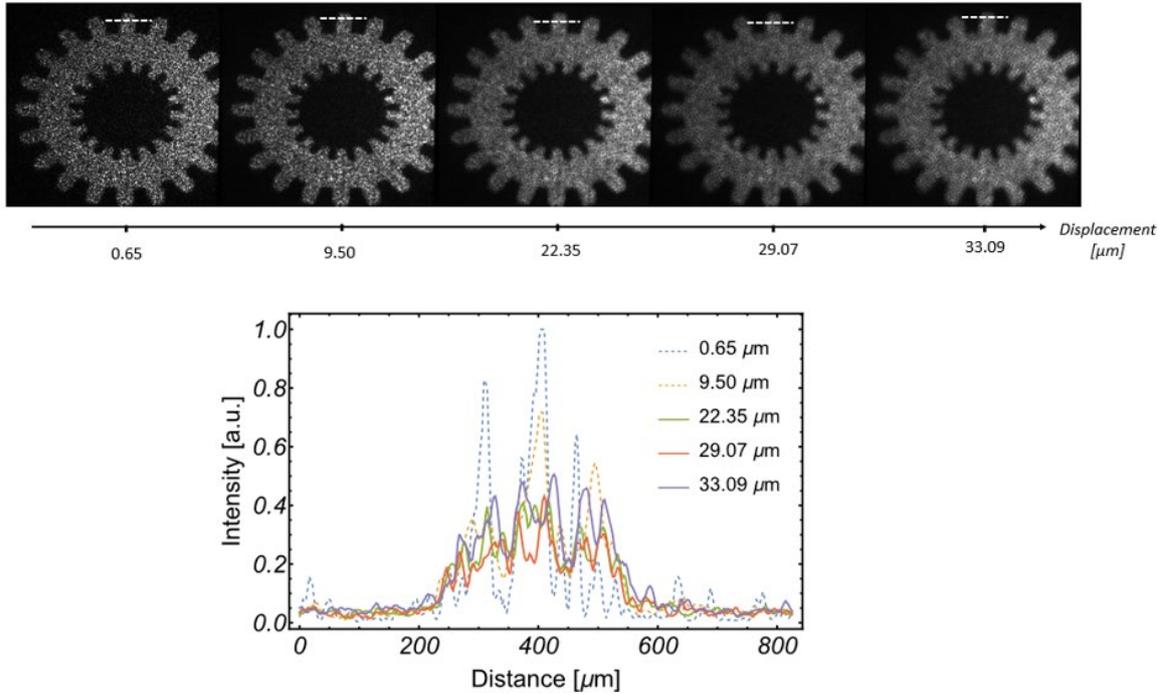

**Supplementary Fig 4.** Top: Examples of accumulated Intensity reconstructions of a gear when 9 different shifts on the axicon phase produce a lateral shift in the image plane. Bottom: Intensity profiles of a gear tooth for different lateral displacements.

**Supplementary Note 4: Axicon shift**

Bessel beams are well known because, unlike Gaussian beams, they are non-diffracting beams whose transverse profile is described by a Bessel function, with a central maximum and concentric rings, and remains unchanged during propagation. There are different approaches to generating Bessel beams[37,38]. For far-field approaches, a ring aperture is used. And for near-field approaches, a conical lens or axicon is used. In this work we are using axicon phases where the mathematical expression to generate a the hologram is the following:

$$\varphi_{PSF}(u,v) = Mod[-k\alpha(n-1)\sqrt{u^2+v^2},\ 2\pi] \qquad \text{(S-2)}$$

Where $k$ is the wave number of the incident beam, $n$ is the refractive index of the axicon, $\alpha$ is the bottom angle of the axicon, and $(u,v)$ is the coordinate with the center of the SLM.

The phase distribution for an axicon with an offset can be described by:

$$\varphi_{PSFshif}(u, v) = Mod[-k\alpha(n - 1)\sqrt{(u - \Delta u)^2 + (v - \Delta v)^2},\ 2\pi] \qquad (S\text{-}3)$$

Where $\Delta u$ and $\Delta v$ are offset of the hologram's center at $(u, v)$, axicon on-axis. Then the vertex coordinate after the offset is $(\Delta u, \Delta v)$, axicon off-axes. Supplementary Fig. 5a. top row, Illustrates the axicon with its vertex center at $(u, v)$ with its corresponding phase and axial propagation. Bottom row, illustrates the axicon with its vertex center at $(\Delta u, \Delta v)$ (off-axis) with its corresponding phase and axial propagation. Supplementary Fig. 5a-b shows the axial light intensity distribution over the propagation, where the axial intensity distribution of the axicon on-axis propagates parallel along the optical axis, while the axial intensity distribution of the axicon off-axis has a small tilt.

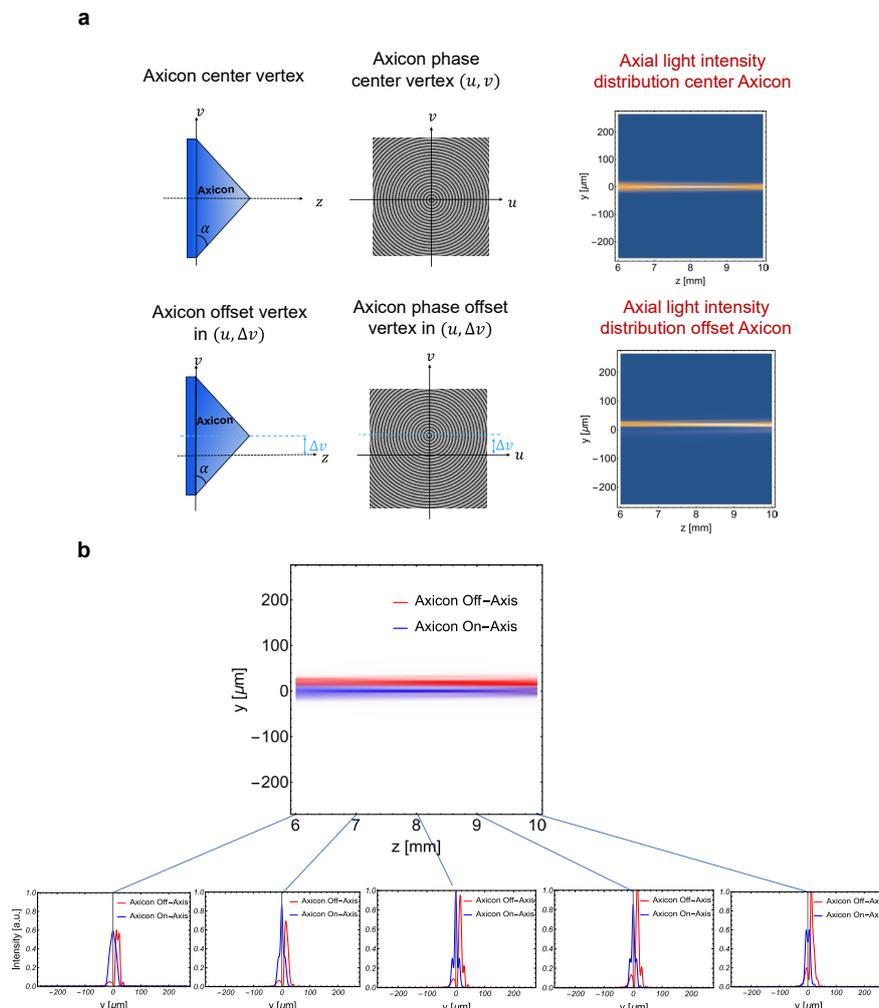

**Supplementary Fig 5. a** Top left: Phase hologram for generating a Bessel beam with the vertex at the center (axicon on-axis, $(u, v)$. Top right: Axial light intensity distribution of a single Bessel beam on-axis. Bottom left: Phase hologram for generating a Bessel beam with the axicon vertex centered at $(u, \Delta v)$. Bottom right: Axial light intensity distribution of a single Bessel beam generated by an off-axis axicon. **b** Combined axial intensities

of the axicon on-axis (blue) and off-axis (red) are shown. Intensity profiles along the propagation direction are displayed

## Supplementary References